# WLAN PERFORMANCE ANALYSIS IBRAHIM GROUP OF INDUSTRIES FAISALABAD PAKISTAN

**Ahmed Mateen**[*]

**Zulafiqar Ali***

**Tasleem Mustafa***

**Abstract**

Now a day's several organizations are moving their LAN foundation towards remote LAN frame work. The purpose for this is extremely straight forward multinational organizations needs their clients surprise about their office surroundings and they additionally need to make wire free environment in their workplaces. Much IT equipment moved on Wireless for instance all in one Pc portable workstations Wireless IP telephones. Another thing is that step by step WLAN innovation moving towards extraordinary effectiveness. In this exploration work Wireless LAN innovation running in Ibrahim Group gathering of commercial enterprises Faisalabad has been investigated in term of their equipment, Wireless signal quality, data transmission, auto channel moving, and security in WLAN system. This examination work required physical proving ground, some WLAN system analyzer (TamoSof throughput) software, hardware point of interest, security testing programming. The investigation displayed in this examination has fill two key needs. One determination is to accept this kind of system interconnection could be broke down utilizing the exploratory models of the two system bits (wired and remote pieces. Second key factor is to determine the security issue in WLAN.

[*] **Lecturer Computer Science, University of Agriculture Faisalabad**

# 1. Introduction

A PC system or information system is information transfers organize that permits PCs to trade information. In PC systems, organized processing gadgets pass information to one another along Information associations. The associations (system connections) between hubs are built up utilizing either link media or remote media. The best-known PC system is the Internet [1]. A remote system is any kind of PC system that uses remote information associations for uniting hubs. Remote systems administration is a technique by which homes, information transfers systems and venture (business) establishments dodge the exorbitant procedure of bringing links into a building, or as an association between different hardware areas. Remote information transfers systems are for the most part executed and regulated utilizing radio correspondence [2]. This execution happens at the physical level (layer) of the OSI model system structure.

## *1.1 What is Throughput?*

In PC innovation, throughput is the measure of work that a PC can do in a given time period. Verifiably, throughput has been a measure of the near viability of expansive business PCs that run numerous projects simultaneously. An early throughput measure was the quantity of group employments finished in a day[3]. Later measures expect a more confused blend of work or concentrate on some specific part of PC operation. While "cost per million directions for every second (MIPS)" gives a premise to looking at the expense of crude figuring after some time or by maker, throughput hypothetically lets you know how much valuable function the MIPS are creating[4].

Another measure of PC profitability is execution, the rate with which one or an arrangement of cluster projects keep running with a sure workload or what number of intuitive client solicitations are being taken care of with what responsiveness. The measure of time between a solitary intelligent client solicitations being entered and getting the application's reaction is known as reaction time.

Clients of in formation transfers gadgets, frameworks fashioners, and specialists into correspondence hypothesis are frequently keen on knowing the normal execution of a framework. From a client point of view, this is frequently stated as either "which gadget will get my information there most adequately for my needs?", or "which gadget will convey the most information per unit cost?". Frameworks planners are regularly intrigued by selecting the best building design or outline imperatives for a framework, which drive its last execution [5]. As a rule, the benchmark of what a framework is prepared to do, or its 'most extreme execution' is the thing that the client or creator is occupied with. While looking at throughput, the term most extreme throughput is every now and again utilized where end-client greatest throughput tests are talked about in subtle element. Greatest throughput is basically synonymous to limit. Four distinct qualities have significance in the setting of "most extreme throughput", utilized as a part of contrasting 'as far as possible' applied execution of different frameworks. They are 'most extreme hypothetical throughput', 'greatest achievable throughput', and 'crest measured throughput' and 'greatest supported throughput'. These speak to diverse amounts and care must be taken that the same definitions are utilized when looking at changed 'most extreme throughput' values [6]. Looking at throughput qualities is additionally subject to every piece conveying the same measure of data. Information pressure can fundamentally skew throughput estimations, including creating qualities more prominent than 100%. On the off chance that the correspondence is interceded by a few connections in arrangement with distinctive piece rates, the most extreme throughput of the general connection is lower than or equivalent to the least piece rate[7]. The most minimal quality connection in the arrangement is alluded to as the bottleneck.

## 1.2 Round-trip time (RTT)

Round-trip time (RTT), conjointly known as round-trip delay, is that the time needed for an indication pulse or packet to travel from a selected supply to a selected destination and back once more. Amid this specific circumstance, the supply is that the portable workstation starting the flag and in this way the goal might be a remote tablet or framework that gets the flag and retransmits it [8]. In broadcast communications, the round-excursion postpone time (RTD) or round-outing time (RTT) is that the length of your time it takes for a sign to be sent and the length of your time it takes for AN affirmation of that flag to be gotten. This point defer so comprises of the proliferation times between the 2 purposes of a sign[9].

With regards to tablet arranges, the flag is normally a parcel, subsequently moreover alluded to online client will check the RTT by exploitation the ping charge. In house innovation, the round-trek postpones time or excursion length is that the time lightweight (and thus any flag) takes to go to a territory test and come. Arrange joins with each a high data measure will truly incredible arrangement (the transmission capacity postpone item) "in funnels" unique convention style [10]. Illustration is that transmission control convention plausibility.

## 2. Related Work

### 2.1 Repeaters

Network repeaters regenerate incoming electrical, wireless or optical signals. With physical media like local area network or Wi-Fi, knowledge transmissions will solely span a restricted distance before the standard of the signal degrades. Repeaters decide to preserve signal integrity and extend the space over that knowledge will safely travel [12].

### 2.2 Hubs

A common place affiliation point for devices in a framework. Focuses are generally used to join parts of a LAN. Middle contains different ports. Exactly when a package gets in contact at one port, it is duplicated to substitute ports with the goal that all segments of the LAN can see all bundles[11].A system extension is programming or equipment that interfaces two or more systems with the goal that they can convey. Individuals with home or little office arranges for the most part utilize a scaffold when they have diverse sorts of systems however they need to trade data or offer records among the greater part of the PCs on those systems[12]. Here's an illustration. Suppose you have two systems: in one, the PCs are joined with links; and in the other, the PCs are associated utilizing remote innovation [13]. The wired PCs can just speak with other wired PCs, and the remote PCs can just correspond with different remote PCs. With a system extension, the greater part of the PCs can correspond with one another.

### 2.3 Switches

A system switch (now and again known as an exchanging center) is a PC organizing gadget that is utilized to interface gadgets together on a PC system by performing a type of bundle exchanging [14]. A switch is viewed as more progressed than a center point on the grounds that a switch will just make an impression on the gadget that needs or demands it, as opposed to TV the same message out of each of its ports. Center points neither give security, or recognizable proof of associated gadgets. This implies messages must be transmitted out of each port of the center, significantly corrupting the productivity of the system [15].

*2.4Routers*

A switch is a gadget that advances data distributes PC frameworks. This makes an overlay internetwork, as a switch is joined with at least two data lines from particular frameworks [16]. Right when a data package comes in one of the lines, the switch scrutinizes the area information in the bundle to choose its authoritative goal. By then, using information as a part of its steering table or steering arrangement, it guides the bundle to the following system on its trip. Switches play out the "development organizing" limits on the Internet. A data package is normally sent beginning with one switch then onto the following through the frameworks that constitute the internetwork until it accomplishes its goal center point. The most unmistakable sort of switches are home and little office switches that essentially pass data, for instance, website pages, email, IM, and recordings between the home PCs and the Internet [17]. A specimen of a switch would be the proprietor's connection or DSL modem, which take up with the Internet through an ISP. More cutting edge switches, for instance, wander switches; join sweeping business or ISP frameworks up to the fit focus switches that forward data at quick along the optical fiber lines of the Internet spine. In spite of the fact that switches are ordinarily devoted equipment gadgets, utilization of programming based switches has developed progressively normal.

*2.5Wireless Access Points*

Remote access focuses (APs or WAPs) are extraordinarily designed gadgets on remote neighborhood (WLANs). Access focuses go about as a focal transmitter and beneficiary of remote radio signs including Wi-Fi. APs are most generally used to bolster open Internet hotspots furthermore on interior business systems to broaden their Wi-Fi sign reach [18]. Access focuses on home or little business systems are little, devoted equipment gadgets including an implicit system connector, reception apparatus, and radio transmitter. Remote switches for home systems incorporate inherent access point usefulness as a component of the gadget. Standalone WAP gadgets likewise exist for both home and business use.

**3. Proposed Work**

*3.1Ibrahim Group of Industries Wireless LAN overview*

In this research work we analyze the wireless local area network performance analysis in term of security, wireless single strength and WIFI link throughput. Here is complete detail of covered area of building.

Table 1.Main Building (Phase-1)

| Basement | 22 | 343.92 Sft |
| --- | --- | --- |
| Ground Floor | 24 | 540.18 Sft |
| First Floor | 24 | 540.18 Sft |
| 2nd Floor | 29 | 012.30 Sft |
| 3rd Floor | 29 | 012.30 Sft |
| 3rd Floor (Mumty) | 03 | 40343 Sft |
| Total Area of Main Building | 103 | 840.01 Sft |

Table 2.Service Building (Phase-2)

| Ground Floor | 14349.00 Sft |
| --- | --- |
| First Floor | 15241.43 Sft |
| 2nd Floor (Mumty) | 29012.30 Sft |
| Total Area of Services Building | 29956.18 Sft |
| Deduction of ABL Bank Branch | 8133.00 Sft |
| Balance Services Building | 21823.18 Sft |

Table 3.External Building (Phase-1)

| Gate office | 1,134.00 Sft |
| --- | --- |
| Visitor Car Parking | 1,323.00 Sft |
| Chiller Room | 2,889.00 Sft |
| Check Post 1 | 64.00 Sft |
| Check Post 2 | 68.00 Sft |
| Total Area of Main Building | 5,478.00 Sft |
| Total Covered area of HO | 139, 274, 19 Sq. Ft |
| False ceiling Height 9'-0" | |
| Total No of Access Points | 7 |
| Total No of WIFI Users | 150 |

## 3.2 Signal Strength Percentage vs. Time

Below mention graph is working of 24 hours which show the signal strength of wireless LAN IFL first floor Hall round about 7 feet distance from AP this reading collected at 4:25 pm. The below mention graph clearly shows that at specified distance round about 7 feet single strength is very good which is round about 70%.Ap used in this research work is Ruckus Zone flex 7431 which is world best AP. Reading started at 4:25 pm and after one hour continuous monitoring it has been observed that signal strength remain at average 70%.In parallel following things also observed which is Total no of frames transmitted count which is 116767.Total number of failed count 4.Total no of retry count 205,Total no of multiple retry count 53,Total no of frame duplicate count 83,Total received fragment count 277821,Ack failure count 324,MulticastRec frame count 10353.

Below mention table shows the values collected during 24 hour working.Table shows that first floor AP works on channel 4 distance from Access point is 10 feet which means that we are very close to our AP data transfer rate of AP is 117 Mbits/sec and its frequency 2427MHz.

### 3.3 Wireless LAN Topology

All networks, whether wired or wireless, have some kind of topology, defined as the network's shape or structure. A wired network's topology is physical and conforms to the layout of cabling in the network. Wireless network topology, on the other hand, is a logical topology conforming to the way the member computers connect and interact with each other, as there is no cable connecting the computers [19].

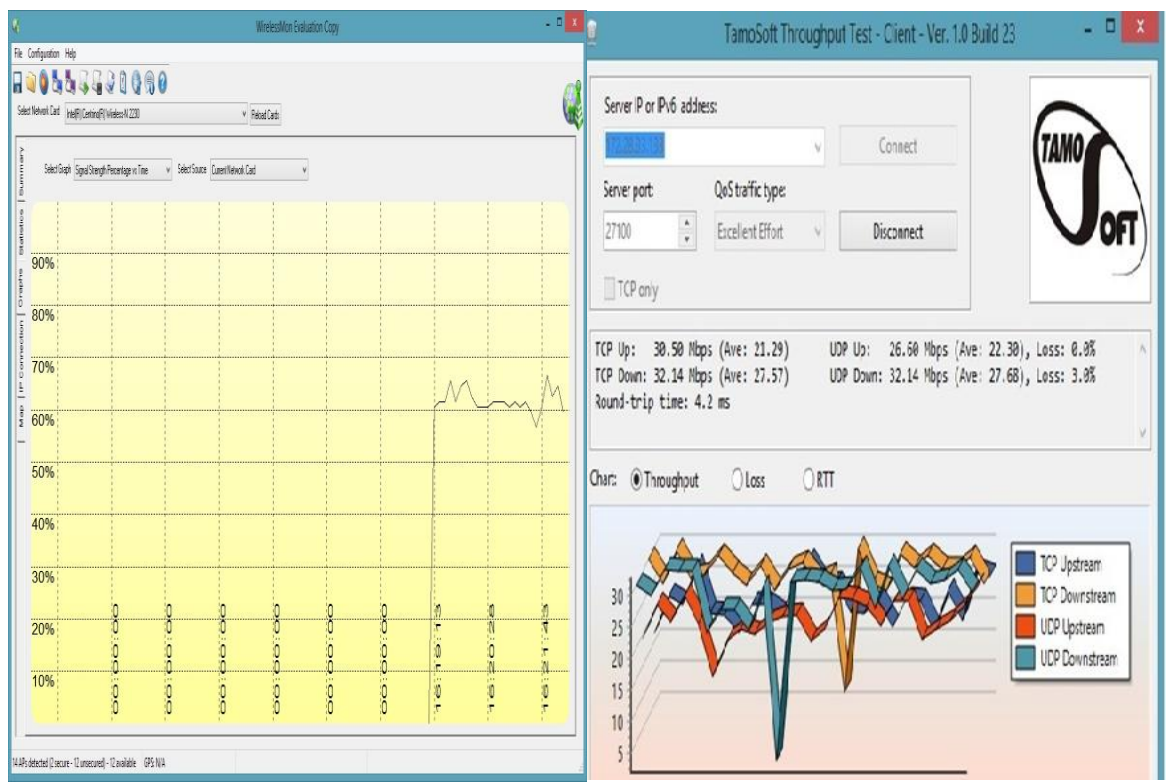

Figure 1.Signal strength vs. Time                    Figure 2.Throuhput

Table 4.Signal strength vs. Time

| Channel Used | Signal Strength | Distance from AP | Speed(Mbits) | Frequency(MHz) |
|---|---|---|---|---|
| 4 | 70% | 10 feet | 117 | 2427 |

*3.4. Wireless Ruckus*

Ruckus is a worldwide Wi-Fi innovation pioneer concentrated on building the up and coming era of Smart Wireless LAN frameworks [20]. The organization makes and markets Smart Wi-Fi items straightforwardly to broadband administrations suppliers and Smart WLAN frameworks in a roundabout way to big business clients. Named a Technology Pioneer by the World Economic Forum, Ruckus Wireless is credited with building up the first Smart Wi-Fi items and innovation that amplify the scope of Wi-Fi signs and consequently adjust to ecological changes [21]. Its clients incorporate substantial telecom administration suppliers, for example, Verizon, AT&T, Time Warner Cable, Deutsche Telekom, and China Telecom. Ruckus Wireless was established in 2004 by William Kish and Victor Storm and is situated in Sunnyvale

*3.5 Throughput at First Floor Hall TCP and UDP*

Below mention graph shows the TCP and UDP downstream and upstream through put at wireless LAN[22].In this graph blue line shows TCP upstream orange line shows TCP downstream, red line shows UDP up stream and sky blue shows the UDP downstream. In this scenario our server is in still position which is connect through a TP link Tl-WN727N wireless dongle to AP and our client is in running position. According to TP link it provides data rate up to 150 Mbps. Now we check through put by moving our client away from server and following thing observed. This reading time stared at 4:00Pm and it is working of 24 hours. Graph clearly shows that when our client moving away from server connected with same IP TCP upstream which means server data send to client is 30Mbps.TCP downstream time is 32Mbps.UDP upstream time is 22Mbps, UDP downstream time is 28.15 Mbps. Tabular data shows clearer picture of our scenario. Below mention Table also shows the Packet loss on Upstream and downstream and also shows the Round trip time.

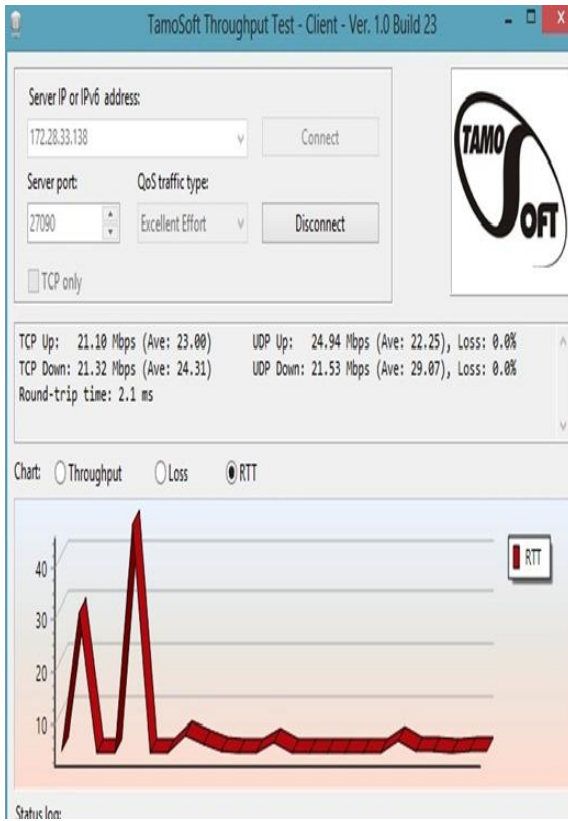 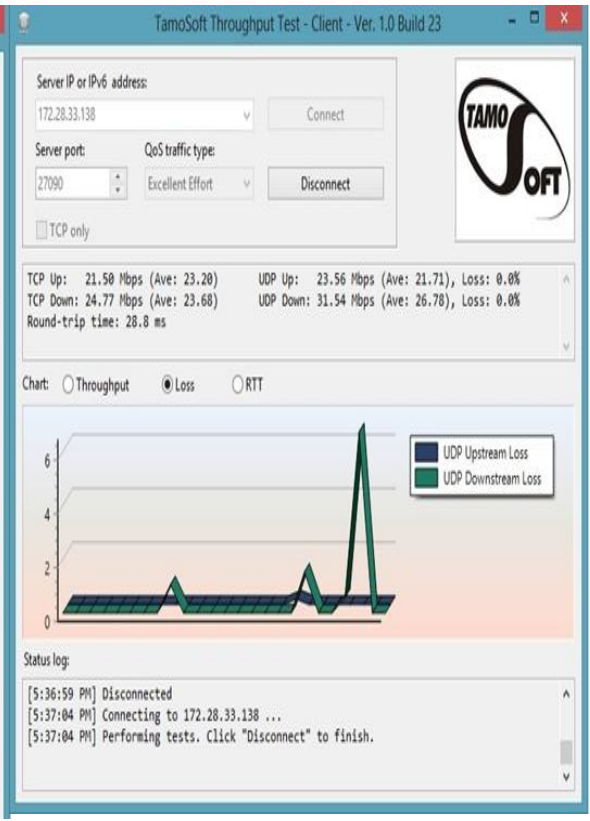

Figure 3.Round Trip Time              Figure 4.(Packet Loss

Table 5.TCP vs. UDP

| TCP Up stream average | TCP Down average | UDp UP average | UDP down Average | Rount Trip Time | Loss Up | Loss Down |
|---|---|---|---|---|---|---|
| 21.29 | 27.59 | 22.66 | 28.15 | 2.1ms | 0.0% | 0.2% |

## 4. Results

TCP and UDP downstream and upstream through put at wireless LAN. In this graph blue line shows TCP upstream orange line shows TCP downstream, red line shows UDP up stream and sky blue shows the UDP downstream. In this scenario our server is in still position which is connect through a TP link Tl-WN727N wireless dongle to AP and our client is in running position. According to TP link it provides data rate up to 150 Mbps. Now we check through put by moving our client away from server and following thing observed. This reading time stared at 4:00Pm and it is working of 24 hours. Graph clearly shows that when our client moving away from server connected with same IP TCP upstream which means server data send to client is 30Mbps.TCP downstream time is 32Mbps.UDP upstream time is 22Mbps, UDP downstream time is 28.15 Mbps. Tabular data shows clearer picture of our scenario.